\documentstyle[12pt]{article}
\def\b{\beta}
\def\t{\theta}

\def\ch{ch}
\def\sh{sh}
\def\ve{\varepsilon}
\def\non{\nonumber}

\begin{document}

\newpage
\pagestyle{empty}
\begin{flushright}
DTP/95-31 \\
hep-th/9506031  \\
June 1995
\end{flushright}
\begin{center}
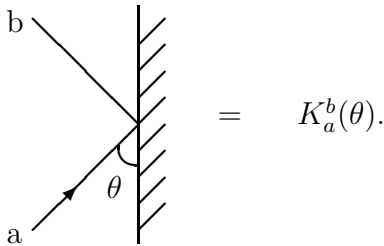
{\Large {\bf Boundary Reflection Matrix \\
for $ade$ Affine Toda Field Theory}}
\end{center}
\begin{center}
{\bf J. D. Kim\footnote{jideog.kim@durham.ac.uk} \footnote{On leave
of absence from Korea Advanced Institute of Science and Technology} } \\
{\it Department of Mathematical Sciences, \\
University of Durham, Durham DH1 3LE, U.K.}
\end{center}
\vspace{2.5cm}

\begin{center}
ABSTRACT
\end{center}

We present a complete set of conjectures
for the exact boundary reflection matrix
for $ade$ affine Toda field theory defined on a half line
with the Neumann boundary condition.

\newpage
\pagestyle{plain}

\section*{I. Introduction}
About a decade ago, studies on the integrable quantum field theory
defined on a half line $(-\infty < x \leq 0)$
was initiated using symmetry principles under the assumption
that the quantum integrability of the model remains intact\cite{Che}.
The boundary Yang-Baxter equation, unitarity relation
for boundary reflection matrix $K_a^b(\t)$
which is conceived to describe the scattering process off a wall
was introduced\cite{Che}.

\begin{picture}(350,130)(-70,-25)
\thicklines
\put(100,95){\line(0,-1){90}}
\put(100,50){\line(-1,1){40}}
\put(100,50){\line(-1,-1){40}}
\put(75,25){\vector(1,1){2}}
\put(50,85){b}
\put(50,5){a}
\put(100,42){\oval(15,15)[b,l]}
\put(88,22){$\t$}
\put(100,80){\line(1,1){10}}
\put(100,70){\line(1,1){10}}
\put(100,60){\line(1,1){10}}
\put(100,50){\line(1,1){10}}
\put(100,40){\line(1,1){10}}
\put(100,30){\line(1,1){10}}
\put(100,20){\line(1,1){10}}
\put(100,10){\line(1,1){10}}
\put(130,50){=}
\put(160,50){$K_a^b(\t)$.}
\put(20,-20){Figure 1. Boundary Reflection Matrix.}
\end{picture}

Recently, the boundary crossing-unitarity relations\cite{GZ}
and the boundary bootstrap equations\cite{FK} have been introduced.
Subsequently, a variety of solutions
of the algebraic equations for affine Toda field theory
has been constructed\cite{GZ,FK,Gh,Sasaki,CDRS,CDR}.
However, a proper interpretation of these solutions in terms of
the Lagrangian quantum field theory had been unknown.

On the other hand, nontrivial boundary potentials which do not
destroy the integrability properties
in the sense that there still exists infinite number of conserved currents
has been determined\cite{GZ,CDRS,CDR,Mac,BCDR,PZ}.
The stability problem of a certain models with boundary potential
has been also discussed\cite{CDR,FR}.

Very recently, we have proposed a formalism\cite{Kim}
to compute a boundary reflection matrix in the framework
of the Lagrangian quantum field theory with a boundary\cite{Sym,DD,BM}.
The idea is to extract the boundary reflection matrix
directly from the two-point correlation function in the coordinate space.
And it has revealed a number of striking features of the perturbative
quantum field theory defined on a half line.

Using this formalism, we determined the exact boundary reflection matrix
for sinh-Gordon model($a_1^{(1)}$ affine Toda theory)
and Bullough-Dodd model($a_2^{(2)}$ affine Toda theory)
with the Neunmann boundary condition
modulo a \lq universal mysterious factor half'\cite{Kim}.
If we assume the strong-weak coupling \lq duality', these solutions
are unique.

Above two models have a particle spectrum with only one mass.
On the other hand, when the theory has a particle spectrum with more
than one mass, each one loop contribution from different types
of Feynman diagrams has a variety of non-meromorphic terms.
We expect actual cancellation of these non-meromorphic terms ought to be
essential for a boundary reflection matrix to have a nice analytic property.

In Ref.\cite{Kim1}, we evaluated one loop boundary reflection matrix for
$d_4^{(1)}$ affine Toda field theory and showed a remarkable cancellation
of non-meromorphic terms among themselves.
This result also enabled us to determine the exact
boundary reflection matrix uniquely under the assumption of
the strong-weak coupling \lq duality'.
It turned out that the boundary reflection matrix has singularities which
can be accounted for by a new type of singularities
of Feynman diagrams for a theory defined on a half line.

In this paper, we present a complete set of conjectures
for the exact boundary reflection matrix
for $ade$ affine Toda field theory defined on a half line
with the Neumann boundary condition.
With this boundary condition, we expect the strong-weak coupling
\lq duality' which is a symmetry of the model defined
on a full line is still effective.

In section II, we review the formalism developed in Ref.\cite{Kim}.
Especially, we give a more informative form of the formulae
given in Ref.\cite{Kim}.
In section III, we present a complete set of conjectures
for the exact boundary reflection matrix
for $ade$ affine Toda field theory with the Neumann boundary condition.
Finally, we make conclusions in section IV.
In appendix, we present one loop result as well as
the complete set of solutions of the boundary bootstrap equations
for $a_3^{(1)}$ theory.

\section*{II. Boundary Reflection Matrix}
The action for affine Toda field theory defined on a half line
($-\infty < x \leq 0$) is given by
\begin{equation}
S(\Phi) = \int_{-\infty}^{0} dx \int_{-\infty}^{\infty} dt
\left ( \frac{1}{2}\partial_{\mu}\phi^{a}\partial^{\mu}\phi^{a}
-\frac{m^{2}}{\b^{2}}\sum_{i=0}^{r}n_{i}e^{\b \alpha_{i} \cdot \Phi}
\right ),
\end{equation}
where
\begin{displaymath}
\alpha_{0} = -\sum_{i=1}^{r}n_{i}\alpha_{i},~~ \mbox{and}~~ n_{0} = 1 .
\end{displaymath}
The field $\phi^{a}$ ($a=1,\cdots,r$) is $a$-th component of the scalar field
$\Phi$,
and $\alpha_{i}$ ($i=1,\cdots,r$) are simple roots of a Lie algebra
$g$ with rank $r$ normalized so that the universal function $B(\b)$
through which the dimensionless coupling
constant $\b$ appears in the $S$-matrix takes the following form:
\begin{equation}
B(\b)=\frac{1}{2\pi}\frac{\b^2}{(1+\b^2/4\pi)}.
\label{Bfunction}
\end{equation}
The $m$ sets the mass scale and the $n_i$s are the so-called Kac labels
which are characteristic integers defined for each Lie algebra.

Here we consider the model with no boundary potential,
which corresponds to the Neumann boundary condition:
$\frac{ \partial \phi^a} {\partial x} =0$ at $x=0$.
This case is believed to be quantum stable in the sense that
the existence of a boundary does not change
the structure of the quantum spectrum determined for the same theory
defined on a full line.

In classical field theory, it is quite clear how we extract
the boundary reflection matrix. It is the coefficient of
reflection term in the classical two-point correlation function
namely it is 1:
\begin{eqnarray}
G_N (t',x';t,x) &=& G(t',x';t,x) + G(t',x';t,-x) \\
             &=& \int \frac{d^2 p}{(2 \pi)^2} \frac{i}{p^2-m_a^2+i \ve}
  e^{-i w (t'-t)}
 (  e^{i k (x'-x)} +  e^{i k (x'+x)} ).  \non
\end{eqnarray}
We may use the $k$-integrated version:
\begin{equation}
G_N (t',x';t,x) = \int \frac{d w}{2 \pi} \frac{1}{2 \bar{k}}
  e^{-i w (t'-t)} (  e^{ i \bar{k} |x'-x| } +  e^{-i \bar{k} (x'+x)} ),
{}~~ \bar{k}=\sqrt{w^2-m_a^2}.
\end{equation}
We find that the unintegrated version is very useful to extract the
asymptotic part of the two-point correlation function far away from the
boundary.

In quantum field theory, it also seems quite natural to extend above
idea in order to extract the quantum boundary reflection matrix directly
from the quantum two-point correlation function.
This idea has been pursued in Ref.\cite{Kim} to extract one loop
boundary reflection matrix.

To compute two-point correlation functions at one loop order,
we follow the idea of the conventional perturbation theory\cite{Sym,DD,BM}.
That is, we generate relevant Feynman diagrams
and then evaluate each of them by using the zero-th order
two-point function for each line occurring in the Feynman diagrams.

At one loop order, there are three types of Feynman diagram contributing to
the two-point correlation function as depicted in figure 2.

\begin{picture}(350,175)(-50,-110)
\thicklines
\put(0,0){\circle{40}}
\put(-20,50){\line(0,-1){100}}
\put(-15,35){a}
\put(-15,-35){a}
\put(25,0){b}
\put(-18,-70){Type I.}
\put(170,0){\circle{40}}
\put(130,0){\line(1,0){20}}
\put(130,50){\line(0,-1){100}}
\put(135,35){a}
\put(135,-35){a}
\put(137,5){b}
\put(195,0){c}
\put(135,-70){Type II.}
\put(300,0){\circle{40}}
\put(300,20){\line(0,1){30}}
\put(300,-20){\line(0,-1){30}}
\put(305,35){a}
\put(305,-35){a}
\put(270,0){b}
\put(325,0){c}
\put(280,-70){Type III.}
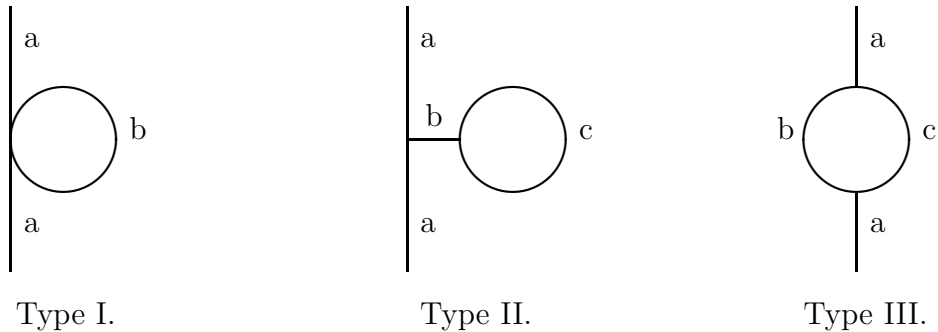
\put(10,-100){Figure 2. Diagrams for the one loop two-point function.}
\end{picture}

For a theory defined on a full line which has translational symmetry
in space and time direction,
Type I, II diagrams have logarithmic infinity
independent of the external energy-momenta
and are the only divergent diagrams in 1+1 dimensions.
This infinity is usually absorbed into
the infinite mass renormalization.
Type III diagrams have finite corrections
depending on the external energy-momenta
and produces a double pole to the two-point correlation function.

The remedy for these double poles is to introduce a counter term
to the original Lagrangian to cancel this term(or to renormalize the mass).
In addition, to maintain the residue of the pole, we have to
introduce wave function renormalization.
Then the renormalized two-point correlation function remains the same
as the tree level one with renormalized mass $m_a$,
whose ratios are the same as the classical value.
This mass renormalization procedure can be generalized
to arbitrary order of loops.

Now let us consider each diagram for a theory defined on a half line.
Type I diagram gives the following contribution:
\begin{equation}
\int_{-\infty}^{0} d x_1  \int_{-\infty}^{\infty} d t_1
 G_N (t,x;t_1,x_1) ~ G_N (t',x';t_1,x_1) ~ G_N (t_1,x_1;t_1,x_1).
\label{TypeI}
\end{equation}
{}From Type II diagram, we can read off the following expression:
\begin{equation}
\int_{-\infty}^{0} d x_1 d x_2 \int_{-\infty}^{\infty} d t_1 d t_2
 G_N (t,x;t_1,x_1) ~ G_N (t',x';t_1,x_1) ~ G_N (t_1,x_1;t_2,x_2)
\end{equation}
\begin{displaymath}
{}~~~~~~~~~~ G_N (t_2,x_2;t_2,x_2).
\end{displaymath}
Type III diagram gives the following contribution:
\begin{equation}
\int_{-\infty}^{0} d x_1  d x_2 \int_{-\infty}^{\infty} d t_1 d t_2
 G_N (t,x;t_1,x_1) ~ G_N (t',x';t_2,x_2) ~ G_N (t_2,x_2;t_1,x_1)
\end{equation}
\begin{displaymath}
G_N (t_2,x_2;t_1,x_1).
\end{displaymath}

After the infinite as well as finite mass renormalization,
the remaining terms coming from type I,II and III diagrams
can be written as follows with different $I_i$ functions\cite{Kim}:
\begin{equation}
\int \frac{dw}{2 \pi} \frac{dk}{2 \pi} \frac{dk'}{2 \pi}
   e^{-iw(t'-t)}  e^{i (kx+k'x')}
 \frac{i}{w^2-k^2-m_a^2+i \ve} \frac{i}{w^2-k'^2-m_a^2+i \ve}
 I_i(w,k,k').
\label{General}
\end{equation}
Contrary to the other terms which resemble those of a full line,
this integral has two spatial momentum integration.

In the asymptotic region far away from the boundary,
these terms can be evaluated up to exponentially damped
term as $x, x'$ go to $-\infty$, yielding the following result
for the elastic boundary reflection matrix $K_a(\t)$ defined
as the coefficient of the reflected term of the two-point
correlation function:
\begin{equation}
\int \frac{dw}{2 \pi} e^{-iw(t'-t)} \frac{1}{2 \bar{k}}
 (  e^{i \bar{k} |x'-x|} +K_a(w)  e^{-i \bar{k} (x'+x)} ),
  ~~ \bar{k}=\sqrt{w^2-m_a^2}.
\end{equation}
$K_a(\t)$ is obtained using $w=m_a \ch\t$.
Here we list each one loop contribution to $K_a(\t)$
from the three types of diagram depicted in figure 2\cite{Kim}:
\begin{eqnarray}
K_a^{(I)}(\t) &=& \frac{1}{4 m_a \sh\t} ( \frac{1}{2 \sqrt{m_a^2
\sh^2\t+m_b^2}}
    +\frac{1}{2 m_b} ) ~C_1 ~S_1,
\label{K-I}  \\
K_a^{(II)}(\t) &=& \frac{1}{4 m_a \sh\t}
  ( \frac{-i}{ (4 m_a^2 \sh^2\t +m_b^2) 2 \sqrt{m_a^2 \sh^2\t+m_c^2}}
    +\frac{-i}{ 2 m_b^2 m_c} ) ~C_2 ~S_2,
\label{K-II}  \\
K_a^{(III)}(\t) &=& \frac{1}{4 m_a \sh\t}
  ( 4 I_3(k_1=0,k_2=\bar{k} )+4 I_3(k_1=\bar{k} ,k_2=0) ) ~ C_3 ~S_3,
\label{K-III}
\end{eqnarray}
where a \lq universal mysterious factor half' is included.
$C_i, S_i$ denote numerical coupling factors and symmetry factors,
respectively. $I_3$ is defined by
\begin{equation}
I_3 \equiv
  \frac{1}{4}
 (\frac{i}{2 \bar{w}_1 (\bar{w}_1-\tilde{w}_1^+) (\bar{w}_1-\tilde{w}_1^-)}
  + \frac{i}{(\tilde{w}_1^+ -\bar{w}_1) (\tilde{w}_1^+ +\bar{w}_1)
        (\tilde{w}_1^+ -\tilde{w}_1^- )  } ),
\end{equation}
where
\begin{eqnarray}
\bar{w}_1=\sqrt{k_1^2+m_b^2}, &
\tilde{w}_1^+ =w+\sqrt{k_2^2+m_c^2}, & \tilde{w}_1^- =w-\sqrt{k_2^2+m_c^2}.
\end{eqnarray}
It should be remarked that this term should be symmetrized with respect
to $m_b, m_c$ with a half.

The expression for a contribution from Type III diagram can be rewritten
in the following form:
\begin{eqnarray}
\label{newK-III}
K_a^{(III)} &=& \frac{i}{4 m_a \sh\t} ~ C_3 ~S_3 \\
 & & (\frac{ \cos\t_{ab}^c }{4 m_a m_b^2 (\ch^2\t-\cos^2\t_{ab}^c) }
  - \frac{ m_a \ch^2\t + m_b \cos \t_{ab}^c }
{2 m_a m_b^2 2\sqrt{m_a^2 \sh^2\t +m_c^2}(\ch^2\t-\cos^2\t_{ab}^c) } \non \\
 &+& \frac{ \cos\t_{ac}^b }{4 m_a m_c^2 (\ch^2\t-\cos^2\t_{ac}^b) }
  - \frac{ m_a \ch^2\t + m_c \cos \t_{ac}^b }
{2 m_a m_c^2 2\sqrt{m_a^2 \sh^2\t +m_b^2}(\ch^2\t-\cos^2\t_{ac}^b) } ), \non
\end{eqnarray}
where $\t_{ab}^c$ is a usual fusion angle defined by
\begin{equation}
\cos\t_{ab}^c= \frac{m_c^2-m_a^2-m_b^2}{2 m_a m_b}.
\end{equation}

Let us note a few interesting points.
Firstly, all the expressions in Eqs.(\ref{K-I},\ref{K-II},\ref{K-III})
have in general non-meromorphic terms when
the theory has a mass spectrum with more than one mass. Cancellation of
these terms is expected to occur for the boundary reflection matrix
to have a nice analytic property. We have verified this non-trivial
cancellation for $d_4^{(1)}$ theory in Ref.\cite{Kim1} and
the result for $a_3^{(1)}$ theory is presented in this appendix.
Secondly, the Feynman diagrams have (simple pole)singularities
which are absent for the theory defined on a full line.
A general study on the analytic property of
the boundary reflection matrix is definitely needed, while that for
the scattering matrix has been extensively done\cite{ELOP}.

Moreover, the position of poles are directly related with fusion angles as in
Eq.(\ref{newK-III}) and less obviously as in Eq.(\ref{K-II}).
Later in the appendix,
we will see a nontrivial cancellation of non-meromorphic terms and
the fact that the new type of singularities accounts for the
singularities of the exact boundary reflection matrix.

\section*{III. The Boundary Reflection Matrix for $ade$ affine Toda theory}
The exact $S$-matrix for integrable quantum field theory defined
on a full line has been conjectured using the symmetry principles
such as Yang-Baxter equation, unitarity, crossing relation,
real analyticity and bootstrap equation\cite{ZZ,AFZ,BCDS,CM,DGZ}.
This program entirely relies on the assumed quantum integrability
of the model as well as the fundamental assumptions
such as strong-weak coupling \lq duality' and \lq minimality'.

In order to determine the exact $S$-matrix uniquely,
Feynman's perturbation theory has been used\cite{BCDS2,BS,CKK,BCKKS,SZ}
and shown to agree well with the conjectured \lq minimal' $S$-matrices.
In perturbation theory, $S$-matrix is extracted
from the four-point correlation function with LSZ reduction formalism.
Especially, the singularity structures were examined in terms of
Landau singularity\cite{ELOP}, of which odd order poles are interpreted
as coming from the intermediate bound states.

In determining the whole set of scattering matrix elements,
it is essentially sufficient to determine the element
for the so-called \lq elementary particle'.
Starting from that element, we can determine all the other elements
using the bootstrap equations\cite{BCDS}. This is also true for the boundary
reflection matrix.
In $a_n^{(1)}$ theory, \lq elementary particle' is the lightest one
corresponding to two end points of the Dynkin digram.
In $d_n^{(1)}$ theory, \lq elementary particles' are those corresponding to
(anti-)spinor representations.
In $e_6^{(1)}$ theory, \lq elementary particles' are the lightest ones
which are conjugate to each other corresponding to two end points
of the Dynkin diagram.
In $e_7^{(1)}$ and $e_8^{(1)}$ theories, it is the lightest one
corresponding to the end point of the longer arm of the Dynkin digram.

\setlength{\unitlength}{0.01cm}
\begin{picture}(1000,250)(-50,50)
\thicklines
\put(292,190){$ \circ$}
\put(310,200){\line( 1, 0){85}}
\put(392,190){$ \circ$}
\put(410,200){\line( 1, 0){85}}
\put(492,190){$ \circ$}
\put(580,190){$ \cdots$}
\put(692,190){$ \circ$}
\put(710,200){\line( 1, 0){85}}
\put(792,190){$ \circ$}
\put(892,190){$ \circ$}
\put(810,200){\line( 1, 0){85}}
\put(292,140){$1$}
\put(392,140){$2$}
\put(780,140){${n-1}$}
\put(892,140){$n$}
\put(300,70){Figure 3. Dynkin diagram for $a_n$.}
\end{picture}

Let us start from $a_n^{(1)}(n \geq 1)$ theory.
The boundary reflection matrix for
the \lq elementary particles' can be coded into the following pyramid
of exponents of the factors $[x]$
which appear in the boundary reflection matrix.

\setlength{\unitlength}{0.01cm}
\begin{picture}(1000,430)(-170,360)
\put(500,700){1}
\put(450,650){1}
\put(550,650){1}
\put(400,600){1}
\put(500,600){1}
\put(600,600){1}
\put(350,550){1}
\put(450,550){1}
\put(550,550){1}
\put(650,550){1}
\put(300,500){1}
\put(400,500){1}
\put(500,500){1}
\put(600,500){1}
\put(700,500){1}
\put(100,400){Figure 4. A pyramid of exponents for $a_n$ theory.}
\end{picture}

It means
\begin{equation}
K_1(\t)=K_n(\t)= \prod_{k=1,step 2}^{2h-3} [k/2],
\end{equation}
where
\begin{equation}
 [ x ] = \frac{ (x-1/2) (x+1/2)} {(x-1/2+B/2) (x+1/2-B/2)},
  ~~~~ (x) = \frac{ \sh( \t /2 + i \pi x /2 h )}
            { \sh( \t /2 - i \pi x /2 h )}.
\label{Blockx}
\end{equation}
{}From these elements of the boundary reflection matrix,
we can in principle determine all the other elements
using the boundary bootstrap equations.

\setlength{\unitlength}{0.01cm}
\begin{picture}(1000,300)(-50,10)
\thicklines
\put(292,190){$ \circ$}
\put(310,200){\line( 1, 0){85}}
\put(392,190){$ \circ$}
\put(410,200){\line( 1, 0){85}}
\put(492,190){$ \circ$}
\put(580,190){$ \cdots$}
\put(692,190){$ \circ$}
\put(710,200){\line( 1, 0){85}}
\put(792,190){$ \circ$}
\put(810,205){\line( 1, 1){70}}
\put(810,195){\line( 1, -1){70}}
\put(875,265){$ \circ$}
\put(875,115){$ \circ$}
\put(260,150){${n-2}$}
\put(370,150){${n-3}$}
\put(690,150){$2$}
\put(780,150){$1$}
\put(910,265){$s$}
\put(910,115){$s' ~or~ \bar{s}$}
\put(300,30){Figure 5. Dynkin diagram for $d_n$.}
\end{picture}

For $d_n^{(1)}(n \geq 2)$ theory,
a pyramid of exponents takes a slightly complicated form.
$d_2^{(1)}$ theory is equal to two copies of sinh-Gordon theory
which is $a_1^{(1)}$ theory
and $d_3^{(1)}$ theory is equal to $a_3^{(1)}$ theory.

\setlength{\unitlength}{0.01cm}
\begin{picture}(1000,400)(-170,380)
\put(500,700){1}
\put(450,650){1}
\put(500,650){1}
\put(550,650){1}
\put(400,600){1}
\put(450,600){1}
\put(500,600){2}
\put(550,600){1}
\put(600,600){1}
\put(350,550){1}
\put(400,550){1}
\put(450,550){2}
\put(500,550){2}
\put(550,550){2}
\put(600,550){1}
\put(650,550){1}
\put(300,500){1}
\put(350,500){1}
\put(400,500){2}
\put(450,500){2}
\put(500,500){3}
\put(550,500){2}
\put(600,500){2}
\put(650,500){1}
\put(700,500){1}
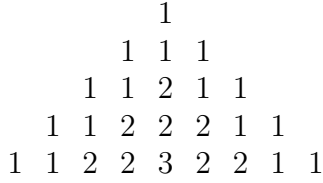
\put(40,420){Figure 6. The first pyramid of exponents for $d_n$ theory.}
\end{picture}

It means
\begin{equation}
K_s(\t)=K_{s'(\bar{s})}(\t)= \prod_{k=1,step 2}^{2h-3} [k/2]^{x_k},
\end{equation}
where $x_k$ are the exponents in sequence from(to) left to(from) right
in figure 6. The rule of the figure 6 is the following.
At odd rows except the apex, prepare two copies of the middle number
and put them to two sites neighbouring to the center, pushing the others
away towards both sides
and increment the original middle number by one unit.
At even rows, do the same thing as for odd rows but leave the middle
number without incrementing.
{}From these elements of the boundary reflection matrix,
we can determine all the other elements.

On the other hand, a pyramid of exponents for lightest particle
corresponding to the end point of the longer arm of the Dynkin diagram
take the following form.

\setlength{\unitlength}{0.01cm}
\begin{picture}(1000,400)(-170,380)
\put(500,700){1}
\put(450,650){1}
\put(500,650){2}
\put(550,650){1}
\put(400,600){1}
\put(450,600){1}
\put(500,600){2}
\put(550,600){1}
\put(600,600){1}
\put(350,550){1}
\put(400,550){1}
\put(450,550){1}
\put(500,550){2}
\put(550,550){1}
\put(600,550){1}
\put(650,550){1}
\put(300,500){1}
\put(350,500){1}
\put(400,500){1}
\put(450,500){1}
\put(500,500){2}
\put(550,500){1}
\put(600,500){1}
\put(650,500){1}
\put(700,500){1}
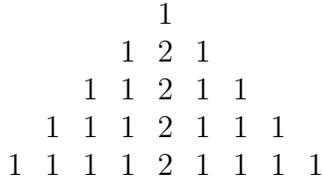
\put(40,420){Figure 7. The second pyramid of exponents for $d_n$ theory.}
\end{picture}

It means
\begin{equation}
K_{n-2}(\t)= \prod_{k=1,step 2}^{2h-3} [k/2]^{x_k},
\end{equation}
where $x_k$ are the exponents in figure 7.
The rule of the figure 7 is that
only the middle number is two except the apex.
{}From these data, we can not determine all the other elements for
each $d_n^{(1)}$ theory.
However, it obviously looks simpler than the elements corresponding
to (anti-)spinor representations.

\setlength{\unitlength}{0.01cm}
\begin{picture}(1000,350)(-180,0)
\thicklines
\put(292,190){$ \circ$}
\put(310,200){\line( 1, 0){85}}
\put(392,190){$ \circ$}
\put(502,207){\line( 0, 1){80}}
\put(492,285){$ \circ$}
\put(410,200){\line( 1, 0){85}}
\put(492,190){$ \circ$}
\put(592,190){$ \circ$}
\put(692,190){$ \circ$}
\put(510,200){\line( 1, 0){85}}
\put(610,200){\line( 1, 0){85}}
\put(292,150){$ 1$}
\put(392,150){$ 3$}
\put(492,150){$ 4$}
\put(592,150){$ 5$}
\put(692,150){$ 6$}
\put(520,300){$ 2$}
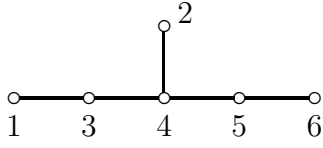
\put(200,50){Figure 8. Dynkin diagram for $e_6$.}
\end{picture}

For $e_n^{(1)}$ theory, we have checked the conjectured boundary reflection
matrices of the \lq elementary particles' by perturbation theory.
Other elements for particles which are not \lq elementary' are determined
using the boundary bootstrap equations.

For $e_6^{(1)}$ theory$(h=12)$, a complete list is
\begin{eqnarray}
K_1(\t) &=& [1/2][3/2][5/2][7/2]^2[9/2]^2[11/2]^2[13/2]^2[15/2]^2[17/2][19/2]
[21/2], \\
K_2(\t) &=& [1/2][3/2][5/2]^2[7/2]^3[9/2]^3[11/2]^3[13/2]^2[15/2]^3[17/2]^2
[19/2][21/2], \non \\
K_3(\t) &=& [1/2][3/2]^2[5/2]^3[7/2]^4[9/2]^4[11/2]^4[13/2]^4[15/2]^3[17/2]^2
[19/2]^2[21/2], \non \\
K_4(\t) &=& [1/2][3/2]^3[5/2]^5[7/2]^6[9/2]^6[11/2]^6[13/2]^5[15/2]^4[17/2]^3
[19/2]^2[21/2],  \non  \\
K_5(\t) &=& K_3(\t), \non \\
K_6(\t) &=& K_1(\t). \non
\end{eqnarray}

\setlength{\unitlength}{0.01cm}
\begin{picture}(1000,350)(-150,0)
\thicklines
\put(292,190){$ \circ$}
\put(310,200){\line( 1, 0){85}}
\put(392,190){$ \circ$}
\put(410,200){\line( 1, 0){85}}
\put(492,190){$ \circ$}
\put(510,200){\line( 1, 0){85}}
\put(592,190){$ \circ$}
\put(602,207){\line( 0, 1){80}}
\put(592,285){$ \circ$}
\put(610,200){\line( 1, 0){85}}
\put(692,190){$ \circ$}
\put(710,200){\line( 1, 0){85}}
\put(792,190){$ \circ$}
\put(792,150){$2$}
\put(292,150){$1$}
\put(392,150){$4$}
\put(492,150){$6$}
\put(592,150){$7$}
\put(692,150){$5$}
\put(620,300){$3$}
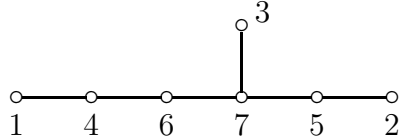
\put(260,50){Figure 9. Dynkin diagram for $e_7$.}
\end{picture}

For $e_7^{(1)}$ theory$(h=18)$,
we report only two elements for technical reasons.
An interested reader should find no difficulty
in producing all the other elements.
A partial list is
\begin{eqnarray}
K_1(\t) &=& [1/2][3/2][5/2][7/2][9/2]^2[11/2]^2[13/2]^2[15/2]^2[17/2]^3
[19/2]^2 \\
 & & [21/2]^2[23/2]^2[25/2]^2[27/2][29/2][31/2][33/2],   \non \\
K_2(\t) &=& [1/2][3/2][5/2][7/2]^2[9/2]^2[11/2]^3[13/2]^3[15/2]^3[17/2]^3
[19/2]^2  \non \\
 & & [21/2]^3[23/2]^3[25/2]^2[27/2]^2[29/2][31/2][33/2]. \non
\end{eqnarray}

\setlength{\unitlength}{0.01cm}
\begin{picture}(1100,350)(-100,0)
\thicklines
\put(292,190){$ \circ$}
\put(310,200){\line( 1, 0){85}}
\put(392,190){$ \circ$}
\put(410,200){\line( 1, 0){85}}
\put(492,190){$ \circ$}
\put(510,200){\line( 1, 0){85}}
\put(592,190){$ \circ$}
\put(610,200){\line( 1, 0){85}}
\put(692,190){$ \circ$}
\put(702,207){\line( 0, 1){80}}
\put(692,285){$ \circ$}
\put(710,200){\line( 1, 0){85}}
\put(792,190){$ \circ$}
\put(810,200){\line( 1, 0){85}}
\put(892,190){$ \circ$}
\put(292,150){$1$}
\put(892,150){$2$}
\put(392,150){$3$}
\put(720,300){$4$}
\put(492,150){$5$}
\put(792,150){$6$}
\put(592,150){$7$}
\put(692,150){$8$}
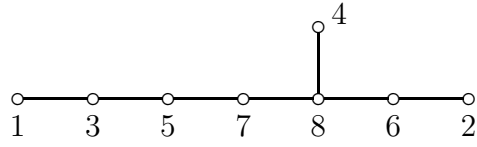
\put(300,50){Figure 10. Dynkin diagram for $e_8$.}
\end{picture}

For $e_8^{(1)}$ theory$(h=30)$, a complete list is
\begin{small}
\begin{eqnarray}
K_1 &=& [1/2][3/2][5/2][7/2][9/2][11/2]^2[13/2]^2[15/2]^2[17/2]^2
[19/2]^3  \\
 & & [21/2]^3[23/2]^3[25/2]^3[27/2]^3[29/2]^4[31/2]^3[33/2]^3[35/2]^3
[37/2]^3[39/2]^3  \non \\
 & & [41/2]^2[43/2]^2[45/2]^2[47/2]^2[49/2][51/2][53/2][55/2]
[57/2], \non \\
K_2 &=& [1/2][3/2][5/2][7/2]^2[9/2]^2[11/2]^3[13/2]^4[15/2]^4[17/2]^5
[19/2]^5 \non \\
 & & [21/2]^4[23/2]^5[25/2]^5[27/2]^5[29/2]^5[31/2]^4[33/2]^5[35/2]^5
[37/2]^4[39/2]^4  \non \\
 & & [41/2]^4[43/2]^4[45/2]^4[47/2]^3[49/2]^2[51/2]^2[53/2]
[55/2][57/2], \non \\
K_3 &=& [1/2][3/2]^2[5/2]^2[7/2]^2[9/2]^3[11/2]^4[13/2]^4[15/2]^4[17/2]^5
[19/2]^6 \non \\
 & & [21/2]^6[23/2]^6[25/2]^6[27/2]^7[29/2]^7[31/2]^6[33/2]^6[35/2]^6
[37/2]^6[39/2]^5  \non \\
 & & [41/2]^4[43/2]^4[45/2]^4[47/2]^3[49/2]^2[51/2]^2[53/2]^2
[55/2]^2[57/2], \non \\
K_4 &=& [1/2][3/2]^2[5/2]^3[7/2]^3[9/2]^4[11/2]^5[13/2]^5[15/2]^6[17/2]^6
[19/2]^7  \non \\
 & & [21/2]^7[23/2]^7[25/2]^7[27/2]^8[29/2]^9[31/2]^8[33/2]^8[35/2]^6
[37/2]^6[39/2]^6  \non \\
 & & [41/2]^5[43/2]^5[45/2]^4[47/2]^4[49/2]^3[51/2]^3[53/2]^2
[55/2]^2[57/2]^2, \non \\
K_5 &=& [1/2][3/2]^2[5/2]^3[7/2]^4[9/2]^6[11/2]^7[13/2]^6[15/2]^7[17/2]^8
[19/2]^9 \non \\
 & & [21/2]^9[23/2]^9[25/2]^{10}[27/2]^{10}[29/2]^9[31/2]^8[33/2]^9
[35/2]^9[37/2]^8[39/2]^7 \non \\
 & & [41/2]^6[43/2]^6[45/2]^5[47/2]^4[49/2]^4[51/2]^4
[53/2]^3[55/2]^2[57/2], \non \\
K_6 &=& [1/2][3/2]^2[5/2]^3[7/2]^4[9/2]^5[11/2]^7[13/2]^8[15/2]^9
[17/2]^{10}[19/2]^{10}  \non \\
 & & [21/2]^{10}[23/2]^{10}[25/2]^{10}[27/2]^{10}[29/2]^{10}
[31/2]^9[33/2]^9[35/2]^9[37/2]^8[39/2]^8 \non \\
 & & [41/2]^7[43/2]^7[45/2]^7[47/2]^5[49/2]^4[51/2]^3[53/2]^2[55/2]^2
[57/2],\non \\
K_7 &=& [1/2][3/2]^2[5/2]^4[7/2]^6[9/2]^7[11/2]^8[13/2]^9[15/2]^{10}
[17/2]^{11}[19/2]^{12}  \non \\
 & & [21/2]^{12}[23/2]^{13}[25/2]^{13}[27/2]^{13}[29/2]^{13}
[31/2]^{12}[33/2]^{12}[35/2]^{11}[37/2]^{10}[39/2]^9 \non \\
 & & [41/2]^8[43/2]^7[45/2]^6[47/2]^5[49/2]^4[51/2]^4[53/2]^3[55/2]^2
[57/2], \non \\
K_8 &=& [1/2]^2[3/2]^4[5/2]^6[7/2]^8[9/2]^9[11/2]^{11}[13/2]^{12}
[15/2]^{13}[17/2]^{14}[19/2]^{15}  \non \\
 & & [21/2]^{15}[23/2]^{15}[25/2]^{16}[27/2]^{16}[29/2]^{17}[31/2]^{16}
[33/2]^{14} [35/2]^{13}[37/2]^{11}[39/2]^{10} \non \\
 & & [41/2]^9[43/2]^8[45/2]^7[47/2]^6[49/2]^5[51/2]^4[53/2]^4[55/2]^3[57/2]^2
[59/2]. \non
\end{eqnarray}
\end{small}

We remark that we have extensive direct proofs for these conjectures
by perturbation theory which are basically case-by-case works.
Parts of them have been already presented in Refs.\cite{Kim,Kim1}
and are presented in the appendix of this paper.
These conjectured boundary reflection matrices are also tested against
various algebraic requirements such as the boundary crossing-unitarity
relations and it always gives consistent results.

\section*{IV. Conclusions}
In this paper, we presented a complete set of conjectures
for the exact boundary reflection matrix for
$ade$ affine Toda field theory defined on a half line with Neumann
boundary condition. These conjectures are based on extensive direct
proofs by perturbation theory and are tested against various algebraic
requirements such as the boundary crossing-unitarity relations
and the boundary boostrap equations.

Surprisingly enough, these solutions have very rich pole structures
in physical strip$(0 \leq Im(\t) <\pi)$.
However, structures of these singularities are explainable in terms of
Feynman diagrams in figure 2 which definitely have no singularity
for the theory defined on a full line and
their positions of poles which are produced by the Feynman diagrams
are related with fusing angles for affine Toda field theory as in
Eq.(\ref{newK-III}).

In the appendix, we presented a detailed computation
for $a_3^{(1)}$ affine Toda field theory up to one loop order
in order to demonstrate a remarkable cancellation of non-meromorphic
terms which are always present for each diagram when the model
has a particle spectrum with more than one mass.
Using this result, we also determined the exact boundary reflection matrix
under the assumption of the strong-weak coupling \lq duality',
which turned out to be \lq non-minimal'. We also presented the complete
set of solutions of the boundary bootstrap equations.

Finally, we remark that a \lq universal mysterious factor half'
which is included in Eqs.(\ref{K-I},\ref{K-II},\ref{K-III})
needs a proper explanation.

\section*{Acknowledgement}
I would like to thank Ed Corrigan, Ryu Sasaki, Harry Braden,
Zheng-Mao Sheng, Hyangsuk Cho, Rachel Rietdijk, Patrick Dorey,
Roberto Tateo and Alistair MacIntyre.
This work was supported by Korea Science and Engineering Foundation
and in part by the University of Durham.

{\bf A note added}: The missing \lq universal mysterious factor half'
is found to be coming from the delta function integral(s)
of the loop spatial momentum(a).
That is, $\int dk~ \delta(2k) =1/2$ instead of 1!

\newpage
\section*{Appendix: $a_3^{(1)}$ affine Toda theory}
We have to fix the normalization of roots so
that the standard $B(\b)$ function takes the form in Eq.(\ref{Bfunction}).

We use the Lagrangian density given by
\begin{eqnarray}
 V(\Phi) &=& 2 m^2 \phi_1 \phi_1^* + 2 m^2 \phi_2 \phi_2
  +i m^2  \b \phi_1 \phi_1 \phi_2 -i m^2 \b \phi_2 \phi_1^* \phi_1^*  \\
 & & - \frac{1}{24} m^2 \b^2 \phi_1 \phi_1 \phi_1 \phi_1
    + \frac{1}{4} m^2 \b^2 \phi_1 \phi_1 \phi_1^* \phi_1^*
    + m^2 \b^2 \phi_1 \phi_1^* \phi_2 \phi_2   \non \\
 & & + \frac{1}{6} m^2 \b^2 \phi_2 \phi_2 \phi_2 \phi_2
  - \frac{1}{24} m^2 \b^2 \phi_1^* \phi_1^* \phi_1^* \phi_1^* +O(\b^3).  \non
\end{eqnarray}
The scattering matrix of this model is given by\cite{AFZ}
\begin{eqnarray}
S_{11}(\t)=S_{33}(\t)=\{1\},&S_{12}(\t)=\{2\},&S_{22}(\t)=\{1\}\{3\},
\end{eqnarray}
\begin{displaymath}
  \{ x \}=\frac{(x-1) (x+1)}{(x-1+B) (x+1-B)}.
\end{displaymath}
Here $B$ is the same function defined in Eq.(\ref{Bfunction}).
For this model, $h=4$ and from now on we set $m=1$.

First we consider the light particle corresponding to $\phi_1$
or its conjugate. It is understood that a suitable choice between a conjugate
pair has to be made depending on a chosen direction of time flow.
There are two possible configurations for Type I diagram.
One is $b=\phi_1$ or its conjugate and
the other is $b=\phi_2$ in the notation of figure 2.
$\phi_1$ loop contribution is the following:
\begin{equation}
K_1(\t)^{(I-1)}= \frac{1}{4 \sqrt{2} \sh\t}
  ( \frac{1}{2 \sqrt{2} \ch\t}+\frac{1}{2 \sqrt{2}})
 \times (\frac{-i}{4} \b^2) \times 4.
\end{equation}
$\phi_2$ loop contribution is the following:
\begin{equation}
K_1(\t)^{(I-2)}= \frac{1}{4 \sqrt{2} \sh\t}
   ( \frac{1}{2 \sqrt{2 \sh^2\t+4} }+\frac{1}{4})
 \times (-i \b^2) \times 1.
\end{equation}

There are no configurations for type II diagram for $a_3^{(1)}$ model.
In fact, this is the case for any $a_n^{(1)}$ theory.

For type III diagram, there exists only one configuration
with $b=\phi_1, c=\phi_2$ symmetrized.
For $b=\phi_1, c=\phi_2$, when $k_1=0, k_2=k$,
\begin{equation}
\bar{w}_1=\sqrt{2},  ~~ \tilde{w}_1^+ = \sqrt{2} \ch\t +\sqrt{2\sh^2\t+4},
                   ~~ \tilde{w}_1^- = \sqrt{2} \ch\t -\sqrt{2\sh^2\t+4},
\end{equation}
and when $ k_1=k, k_2=0 $,
\begin{equation}
\bar{w}_1=\sqrt{2\sh^2\t+2}, ~~ \tilde{w}_1^+ = \sqrt{2} \ch\t+2,
                             ~~ \tilde{w}_1^- =\sqrt{2} \ch\t-2.
\end{equation}
For $b=\phi_2, c=\phi_1$, when $ k_1=0, k_2=k $,
\begin{equation}
\bar{w}_1=2,  ~~ \tilde{w}_1^+ = \sqrt{2} \ch\t +\sqrt{2\sh^2\t+2},
                   ~~ \tilde{w}_1^- = \sqrt{2} \ch\t -\sqrt{2\sh^2\t+2},
\end{equation}
and when $ k_1=k, k_2=0 $,
\begin{equation}
\bar{w}_1=\sqrt{2\sh^2\t+4}, ~~ \tilde{w}_1^+ = \sqrt{2} \ch\t+\sqrt{2},
                             ~~ \tilde{w}_1^- =\sqrt{2} \ch\t-\sqrt{2}.
\end{equation}
The result for Type III diagram can be obtained by inserting above data into
Eq.(\ref{K-III}):
\begin{equation}
K_1(\t)^{(III)}= \frac{1}{4 \sqrt{2} \sh\t}
   ( -\frac{i}{8 \sqrt{2} \ch\t } - \frac{i}{8 \sqrt{2} \sqrt{\sh^2\t+2} }
  +\frac{i}{16 ( \sqrt{2} \ch\t+1) } )
 \times (-\b^2) \times 4.
\end{equation}

Adding the above contributions as well as the classical
value 1, boundary reflection matrix for the light particle is given by
\begin{equation}
K_1(\t)=1+ \frac{i \b^2}{16}
 ( \frac{\sh\t}{\ch\t+1/\sqrt{2}}-\frac{\sh\t}{\ch\t-1}) +O(\b^4).
\end{equation}
The unwanted non-meromorphic terms exactly cancel out.

Now we consider the heavy particle corresponding to $\phi_2$
which are self conjugate.
There are two possible configurations for Type I diagram.
One is $b=\phi_1$, the other is $b=\phi_2$ in the notation of figure 2.
$\phi_2$ loop contribution is the following:
\begin{equation}
K_2(\t)^{(I-1)}= \frac{1}{8 \sh\t} ( \frac{1}{4 \ch\t}+\frac{1}{4})
    \times (\frac{-i}{6} \b^2) \times 12.
\end{equation}
$\phi_1$ loop contribution is the following:
\begin{equation}
K_2(\t)^{(I-2)}= \frac{1}{8 \sh\t}
  ( \frac{1}{2 \sqrt{ 4 \sh^2 \t+2} }+\frac{1}{2 \sqrt{2}})
  \times (-i \b^2) \times 2.
\end{equation}

There is no type II diagram for the heavy particle, either.

For type III diagram, there is single configuration with $b=\phi_1, c=\phi_1$.
When $k_1=0, k_2=k$,
\begin{equation}
\bar{w}_1=\sqrt{2},  ~~ \tilde{w}_1^+ = 2 \ch\t +\sqrt{4\sh^2\t+2},
                   ~~ \tilde{w}_1^- = 2 \ch\t -\sqrt{4\sh^2\t+2},
\end{equation}
and when $ k_1=k, k_2=0$,
\begin{equation}
\bar{w}_1=\sqrt{4\sh^2\t+2}, ~~ \tilde{w}_1^+ = 2 \ch\t+\sqrt{2} ,
                             ~~ \tilde{w}_1^- =2 \ch\t-\sqrt{2}.
\end{equation}
The result for Type III diagram can be obtained by inserting above data
into Eq.(\ref{K-III}):
\begin{equation}
K_2(\t)^{(III)}= \frac{1}{8 \sh\t}
   ( \frac{-i}{8 \sqrt{2} (\sqrt{2} \ch\t-1) }
       +\frac{i}{8 \sqrt{2} (\sqrt{2} \ch\t+1) }
       - \frac{i}{4 \sqrt{2} \sqrt{2 \sh^2\t+1} } )
\end{equation}
\begin{displaymath}
 \times (- \b^2) \times 4.
\end{displaymath}

Adding the above contributions as well as the classical value 1,
boundary reflection matrix for the heavy particle is given by
\begin{equation}
K_2(\t)=1+ \frac{i \b^2}{16}( \frac{\sh\t}{\ch\t} -\frac{\sh\t}{\ch\t-1}
  - \frac{\sh\t}{\ch\t-1/\sqrt{2}} +\frac{\sh\t}{\ch\t+1/\sqrt{2}} )
 +O(\b^4).
\end{equation}
The unwanted non-meromorphic terms exactly cancel out once again.

On the other hand, there are two \lq minimal' boundary reflection matrices
are known for $a_3^{(1)}$ model\cite{FK,Sasaki}.
None of these agrees with the perturbative result.

We have checked that this boundary reflection matrix at one loop order
by perturbation theory satisfies the boundary crossing-unitarity relations
as well as the boundary bootstrap equations:
\begin{equation}
\begin{array}{l}
K_1(\t) ~ K_1(\t-i \pi) = S_{11}(2 \t),
 ~~~  K_2(\t) ~ K_2(\t-i \pi) = S_{22}(2 \t),  \\
K_2(\t) = K_1(\t+i \pi/4) ~K_1(\t-i \pi/4) ~ S_{11}(2 \t),  \non \\
K_1(\t)=K_3(\t).   \non
\end{array}
\end{equation}
In one loop checks, the following identity is useful:
\begin{equation}
\frac{(x+B/2)}{(x)} =1 +
 \frac{i \pi B}{2 h} \frac{\sh\t}{\ch\t-\cos(x\pi /h)} +O(B^2).
\end{equation}
The exact boundary reflection matrix is determined uniquely
if we assume the strong-weak coupling \lq duality':
\begin{eqnarray}
K_1(\t) &=& [1/2] [3/2] [5/2], \\
K_2(\t) &=& [1/2] [3/2]^2 [5/2].  \non
\end{eqnarray}

On the other hand, the most general solution can be written in the
following form under the assumption of the strong-weak coupling
\lq duality':
\begin{equation}
K_1(\t)= [1/2]^{a_1} [3/2]^{b_1} [5/2]^{c_1} [7/2]^{d_1},
\end{equation}
\begin{displaymath}
K_2(\t)= [1/2]^{a_2} [3/2]^{b_2} [5/2]^{c_2} [7/2]^{d_2}.
\end{displaymath}
Inserting the above into the boundary bootstrap equations, we can obtain
linear algebraic relations among the exponents. Solving this system of
equations yields
\begin{equation}
\begin{array}{llll}
a_1 = free,       & b_1 = free,     & c_1 =b_1,        & d_1 =a_1-1,  \non \\
a_2 = -a_1+b_1+1, & b_2 = a_1 +b_1, & c_2 =a_1+ b_1-1, & d_2 =-a_1+ b_1. \non
\end{array}
\end{equation}

\end{document}